\documentclass[conference]{IEEEtran}
\IEEEoverridecommandlockouts

\usepackage{cite}
\usepackage{amsmath,amssymb,amsfonts}
\usepackage{algorithmic}
\usepackage{graphicx}
\usepackage{textcomp}
\usepackage{xcolor}
\def\BibTeX{{\rm B\kern-.05em{\sc i\kern-.025em b}\kern-.08em
    T\kern-.1667em\lower.7ex\hbox{E}\kern-.125emX}}
\begin{document}

\title{National Data Platform's Education Hub\\
\thanks{NDP is funded by the US National Science Foundation award \#2230081. The Fire-Ready Forests Data Challenge was funded by US National Science Foundation award \#2341120.}
}

\author{
\IEEEauthorblockN{Pedro Ramonetti}
\IEEEauthorblockA{\textit{San Diego Supercomputer Center} \\
\textit{University of California San Diego}\\
La Jolla, CA, USA \\
pramonettivega@ucsd.edu}
\and
\IEEEauthorblockN{Melissa Floca}
\IEEEauthorblockA{\textit{San Diego Supercomputer Center} \\
\textit{University of California San Diego}\\
La Jolla, CA, USA \\
mfloca@ucsd.edu}
\and
\IEEEauthorblockN{Kate O'Laughlin}
\IEEEauthorblockA{\textit{San Diego Supercomputer Center} \\
\textit{University of California San Diego}\\
La Jolla, CA, USA \\
kbolaughlin@ucsd.edu}
\and
\IEEEauthorblockN{\hspace{1cm}Amarnath Gupta}
\IEEEauthorblockA{\textit{\hspace{1cm}San Diego Supercomputer Center} \\
\textit{\hspace{1cm}University of California San Diego}\\
\hspace{1cm}La Jolla, CA, USA \\
\hspace{1cm}a1gupta@ucsd.edu} 
\and 
\IEEEauthorblockN{\hspace{1cm}Manish Parashar}
\IEEEauthorblockA{\textit{\hspace{1cm}University of Utah} \\
\hspace{1cm}Salt Lake City, UT, USA \\
\hspace{1cm}manish.parashar@utah.edu}
\and
\IEEEauthorblockN{\hspace{1cm}Ilkay Altintas}
\IEEEauthorblockA{\hspace{1cm}\textit{San Diego Supercomputer Center} \\
\textit{\hspace{1cm}University of California San Diego}\\
\hspace{1cm}La Jolla, CA, USA \\
\hspace{1cm}ialtintas@ucsd.edu}
}

\maketitle
\begin{abstract} 
As demand for AI literacy and data science education grows, there is a critical need for infrastructure that bridges the gap between research data, computational resources, and educational experiences. To address this gap, we developed a first-of-its-kind Education Hub within the National Data Platform. This hub enables seamless connections between collaborative research workspaces, classroom environments, and data challenge settings. Early use cases demonstrate the effectiveness of the platform in supporting complex and resource-intensive educational activities. Ongoing efforts aim to enhance the user experience and expand adoption by educators and learners alike.
\end{abstract}


\section{Introduction}\label{INTRODUCTION}
The rapid expansion of artificial intelligence (AI) and data science has led to a growing demand for accessible, hands-on educational opportunities that prepare students and professionals to work with real-world data at scale. However, a persistent gap exists between advanced data and compute resources used in research environments and tools commonly available in educational settings.

Although national research platforms offer robust infrastructure for data-driven discovery, their potential to support education remains underdeveloped.

Educators often face barriers when attempting to integrate large-scale data resources or high-performance computing (HPC) into coursework. These challenges include complex access protocols, limited technical expertise, and a lack of educational tools embedded in the research infrastructure. As a result, students are frequently restricted to simplified or synthetic datasets and isolated learning environments that do not reflect the collaborative, data-intensive workflows of modern research and industry.

\section{The NDP Education Hub: Vision and Design}\label{VISION}

The Education Hub is an educational driven space designed as part of the National Data Platform (NDP) \cite{ndp1,ndp2} an extensible ecosystem built on national cyberinfrastructure to address critical gaps in foundational data services and infrastructure. This ecosystem enables
    \begin{itemize}
        \item \textbf{Federation of siloed data repositories} into a unified platform for discovery, access, and use;
        \item \textbf{Integration with advanced computing environments} for scalable data analysis; and
        \item \textbf{Standardized, customizable near-data services} for data ingestion, indexing, curation, and analysis.
    \end{itemize} 

\begin{figure}[h!]
    \centering
    \includegraphics[width=\linewidth]{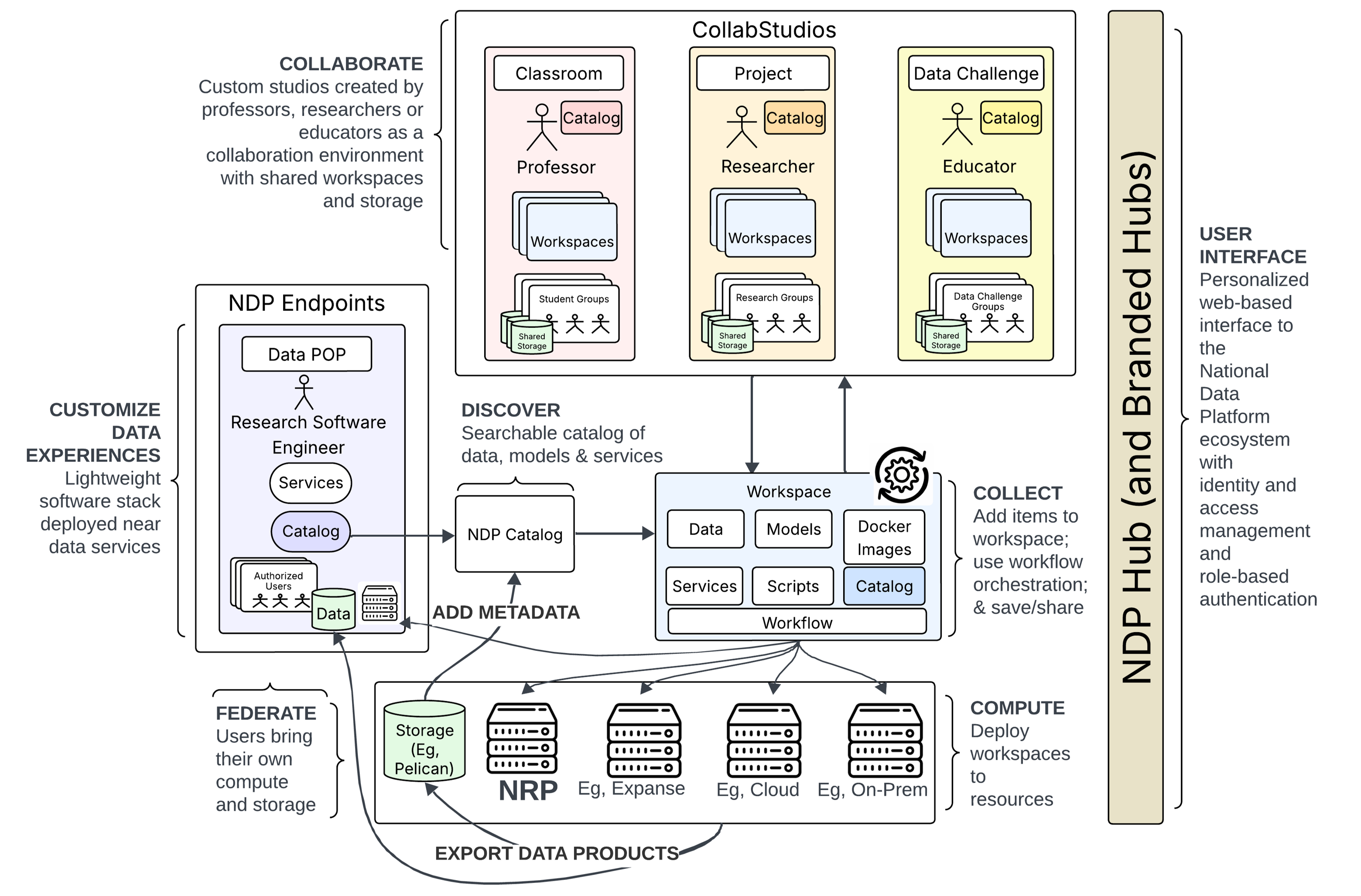}
    \caption{NDP's Architecture}
    \label{fig:sprints}
\end{figure}

The Education Hub currently supports two types of educational spaces, each designed to serve distinct learning and engagement goals.

\begin{itemize}
\item \textbf{Classrooms} are private spaces designed to deliver resource-intensive courses, particularly those focusing on AI and HPC.
\item \textbf{Data Challenges} are open spaces focused on solving real-world problems. Participants are provided with complex datasets and example workflows to guide the development of a solution.
\end{itemize}

At the core of the Education Hub experience are two key components: workspaces support for collaborative learning.

\subsection{Workspaces}

Both classrooms and data challenges in the Education Hub rely on the integration of workspaces. These workspaces, also called modules within the Education Hub, enable educators and students to bundle datasets, computing and data services, open source code (e.g., from GitHub) and AI/ML models (e.g., from Hugging Face) into a single, deployable unit. Each workspace is designed for seamless deployment on HPC infrastructure, including national resources, cloud allocations, or on-prem systems, primarily through JupyterHub. A user-friendly creation wizard further supports this process by guiding users through a detailed workspace configuration and allowing the inclusion of supplementary resources such as files and external links.

Educators can leverage workspaces in a variety of ways, including the following use cases.

\begin{itemize}
    \item \textbf{Learning/Training Unit:} Workspaces serve as the foundation for a learning unit, typically structured around an instructional Jupyter notebook.
    \item \textbf{Assignments:} Instructors can use workspaces as templates for specific assignments within a learning unit.
    \item \textbf{Project Assessment:} Students can submit final projects as self-contained, ready-to-deploy workspaces.
\end{itemize}

\subsection{Collaboration Space}

The Education Hub is built as a collaborative environment where learners work in groups, whether in data challenges or classroom settings, to foster teamwork and shared problem solving.

Each group can modify their assigned workspaces by adding datasets, source code, and other relevant resources. Within NDP’s JupyterHub, groups are also given access to a shared storage directory, which supports file sharing and the preservation of intermediate results.

\section{Significance and Novelty}\label{SIG&NOV}

The Education Hub offers a novel educational resource by integrating real-world datasets with the computing infrastructure required to develop complex workflows. It brings together data, computing power, and multiple services into a unified environment that is accessible to both educators and learners.

Early use cases have demonstrated the effectiveness of the Education Hub include: 

\begin{itemize}
    \item a graduate course in Data Science and Engineering at UC San Diego, which used an NDP Classroom to guide students through final projects involving knowledge graphs. This included access to Neo4j databases hosted on NDP and customized environments, delivered through workspaces, preloaded with all necessary dependencies.
    \item a Data Challenge, held from February to April 2025, which tasked participants with developing a machine learning workflow to extract tree metrics from Terrestrial Laser Scanning (TLS) data. The challenge provided instructional workspaces to introduce students to different data formats, and participants submitted their final solutions as deployable workspaces.
\end{itemize}

In both cases, the students worked with tens of gigabytes of data and were able to launch ready-to-use Jupyter notebooks on a Kubernetes cluster provided by the National Research Platform, capabilities that would have been difficult or impossible to achieve using their own personal machines. Without the Education Hub, these activities would have required significant technical expertise just to set up.

Although some unresolved technical requirements and challenges remain, these use cases highlight how the Education Hub significantly reduces the barrier to entry for working with large-scale data and advanced computing environments in education.

\section{Future Work}\label{Lessons-Learned}

Our initial use cases have provided valuable insights that continue to inform the evolution of the platform and support broader user adoption.

Looking ahead, our efforts will focus on enhancing the educator experience. This includes potential integration with Learning Management Systems (LMS) and the development of built-in grading tools.

We also aim to expand support for additional working environments beyond JupyterHub, with a particular focus on reducing the gap between user skill levels and the technical complexity of HPC infrastructure.

Lastly, we will continue to develop onboarding materials, sample projects, and instructional workspaces to make it easier for educators and learners to adopt and succeed with the Education Hub.

This work is an open invitation for educators to adopt the NDP Education Hub to enrich their classrooms with hands-on AI and HPC instruction. We encourage the broader educational community to host their courses and data challenges on the platform, contributing to a growing collection of instructional materials, and helping shape the future of the Education Hub through real-world use and feedback.

\section*{Acknowledgment}
NDP is funded by the US National Science Foundation award \#2333609. The Fire-Ready Forests Data Challenge was funded by US National Science Foundation award \#2341120.

\bibliographystyle{IEEEtran}
\bibliography{refs}

\end{document}